\documentclass[twocolumn]{svmult}

\usepackage{a4wide}
\usepackage{type1cm}        
\usepackage{makeidx}         
\usepackage{graphicx}        
\RequirePackage{subfig}
\RequirePackage{bm}
\usepackage{multicol}        
\usepackage[bottom]{footmisc}

\usepackage{newtxtext}       %
\usepackage{newtxmath}       

\usepackage{amsmath}
\usepackage{url}
\usepackage{ulem}

\usepackage{xcolor, soul}
\sethlcolor{white}

\graphicspath{{./fig/}}

\makeindex             

\begin{document}

\title*{Numerical Methods for Flow in Fractured Porous Media}
\author{Formaggia, Luca and Scotti, Anna and Fumagalli, Alessio}
\authorrunning{Formaggia, L. and Scotti, A. and Fumagalli, A.}
\institute{Luca Formaggia \at Politecnico di Milano, piazza Leonardo da Vinci
32, 20133 Milan, Italy \email{luca.formaggia@polimi.it}
\and Anna Scotti \at Politecnico di Milano, piazza Leonardo da Vinci
32, 20133 Milan, Italy \email{anna.scotti@polimi.it}
\and Alessio Fumagalli \at Politecnico di Milano, piazza Leonardo da Vinci
32, 20133 Milan, Italy \email{alessio.fumagalli@polimi.it}}
%
%
\maketitle

\section*{Definition}
\begin{itemize}
    \item porous medium: A material containing voids (pores), whose size is small compared to the size of the sample.  The pores are typically filled with a fluid. A porous medium is characterized by macroscopic properties which are obtained with an averaging procedure.
    \item fracture: a break in a material with a small thickness compared to its global extension.
    \item fracture network: a network composed by several intersecting fractures.
\end{itemize}

\section{Introduction}

Fractures are ubiquitous in \hl{porous media}. Here, with the term
\hl{fracture} we
denote a void in the porous material that has the following
characteristics.

\begin{figure}[tbp]
    \centering
    \includegraphics[width=0.4\textwidth]{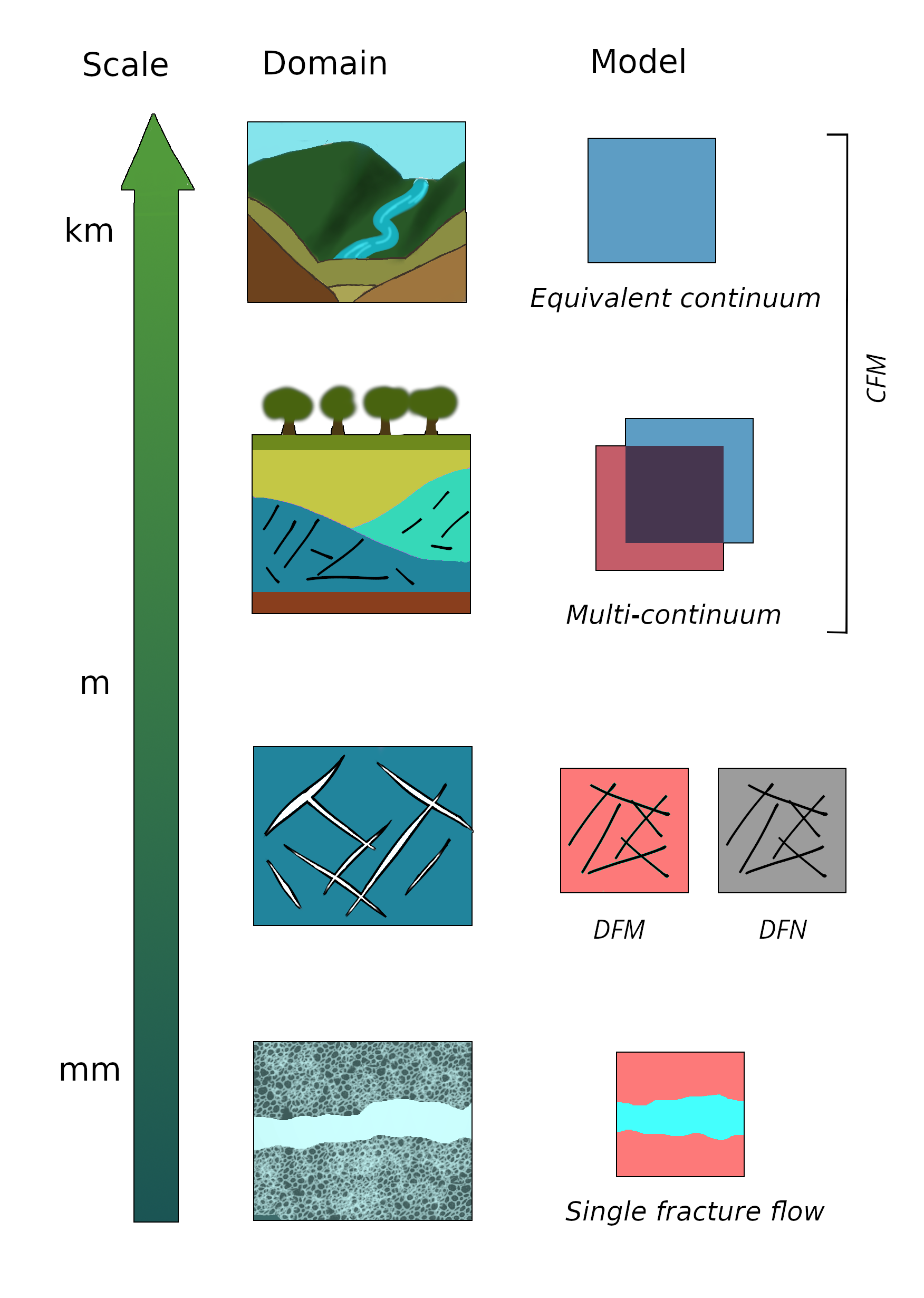}
    \caption{Choice of the numerical model for domains at different space
    scales.}%
    \label{fig:scale}
\end{figure}

\textit{i)} One of its dimensions, the
\hl{aperture}, is orders of magnitude smaller than the other
dimensions and the size of the
domain of interest, but still large compared to pore size.
We will indicate with {extension} the size of a fracture in the directions orthogonal to the
aperture. The extension of
fractures in a network has a distribution that is usually assumed to
be governed  by a power law, which implies the presence of a large variation
of space scales. See Fig~\ref{fig:scale}.

\textit{ii)} Fractures may be either open or infilled by
a porous medium, whose
physical characteristics may be strongly different from those of the
surrounding material.

\textit{iii)} Fractures usually form networks, often highly connected. An
example of \hl{fracture network} is depicted in
Fig~\ref{fig:sotra}.

With the previous definitions one  can consider as fractures,
 depending on the scale of interest,  different objects such as: tectonic faults at the scale of
sedimentary basins, cracks in glaciers, as well as fractures in concrete or in rocks.

When they are empty or filled with highly permeable materials, fractures may provide a preferential path to fluid flow,
but in some cases the deposits inside fractures can become nearly impermeable.
We refer to the latter situation as {{blocking} fracture}.

The presence of fractures greatly alters the macroscopic properties of the material
in a complex way, in particular its mechanical and flow characteristics.
We will be here concerned with the second aspect,
and specifically on the mathematical modelling and computing techniques that may
be adopted in the presence of fractured porous media. For a more general
treatment of fractures in porous media the reader may refer to the Book of P.M.
Adler et al.~\cite{Adler2012}.

{The irregular spatial distribution of fractures and the
presence of multiple scales} makes it difficult and often impossible to account
for fractures by deriving effective upscaled parameters, permeability for
instance, by volume averaging or homogeneization techniques. Indeed, such
procedures assume a strong separation of scales.  Therefore, methods
have been developed that model fractures explicitly, at least those crucial
for the flow.  These models are based on the assumption that in the porous medium
flow is governed by Darcy's law, and
often a similar model is adopted for the flow taking place in the fractures as well.

We recall the main characteristic of the \hl{Darcy's model}, considering,
for simplicity, just the case of single-phase flow. In this mathematical
framework the two main variables are the \hl{pressure} $p$
and the macroscopic velocity field  $\bm{u}$, also called \hl{Darcy's velocity}.
The two quantities are related by  Darcy's law,
\begin{subequations}\label{eq:darcy_cont}
\begin{equation}\label{eq:Darcy}
    \mu \bm{K}^{-1} \bm{u} + \nabla p = \bm{0} \quad \text{in } \Omega,
\end{equation}
where $\Omega\subset\mathbb{R}^d$ represents the domain occupied by the porous
material and $\nabla$ indicates the gradient. In the case where gravity effect are
relevant, equation~\eqref{eq:darcy_cont} may be modified by replacing the
pressure term with $p-\rho gz$, where $\rho$ is the fluid density, $g$ the magnitude of
the gravity acceleration and $z$ is the vertical coordinate pointing upwards from
the Earth surface. The main hypotheses behind the model are that fluid velocity
is small, so we can neglect inertial effects, and the main model parameters are:
$\mu$, the fluid {viscosity}, and $\bm{K}$, the
\hl{permeability} tensor of the porous medium,  which is a symmetric and positive definite tensor.
Permeability may be heterogeneous in space and often with high
variations.

The second equation expresses {continuity of mass} by the following differential
equation,
\begin{equation}\label{eq:cont}
    c\phi {\partial_t p} + \nabla\cdot\bm{u}=q \quad \text{in } \Omega,
\end{equation}
\end{subequations}
where $\nabla\cdot$ is the divergence operator, $q$ a source/sink term, {$c$
accounts for the medium and fluid compressibility and $\phi$ is the porosity}.
Sometimes one is interested in the steady state solution {or the compressibility
can be neglected}, in which case $c\phi\partial_t p = 0$.

Equations~\eqref{eq:darcy_cont} form a system of
partial differential equations which, complemented by
appropriate boundary and initial conditions, allows to describe the evolution of
$(\bm{u}, p)$ in the porous medium.

\section{Numerical models for fractured porous media}

Numerical models for fractured porous media may be subdivided into two main categories:
\hl{Continuum Fracture Models} (CFMs) and \hl{Discrete Fracture Models
} (DFMs).

\subsection{Continuum Fracture Models}
    CFMs are early models
    introduced in the 1960's~\cite{Warren1963}, later justified mathematically
    in~\cite{Arbogast1990}, and currently implemented in many industrial
    software. They assume a highly permeable and interconnected fracture network so that
    it can be modeled as a continuum superimposed to that of the porous medium.

    A commonly adopted method is the
    \hl{dual-porosity/dual-permeability} scheme. It assumes that at each
    point of the domain $\Omega$ we may use the Darcy's
    equations for flow in the fractures and in the porous medium, respectively, with a term representing
    the interchange of mass between them.  A basic model of this type may be written
    as:%
    \begin{subequations}\label{eq:cfm}
    \begin{gather}
        \begin{aligned}
            &c_m\phi_m \partial_t p_m+\nabla\cdot\bm{u}_m-\alpha(p_m-p_f)=q_m \\
            &c_f\phi_f \partial_t p_f+\nabla\cdot\bm{u}_f+\alpha(p_m-p_f)=q_f
        \end{aligned}
        \quad \text{in } \Omega,
    \end{gather}
    where suffixes $m$ and $f$ refer to quantities related to the porous medium and
    fractures, respectively. The term $\alpha(p_m-p_f)$ represents the mass exchange
    between the two components, with $\alpha$ a rate parameter. The Darcy's
    velocities $\bm{u}_m$ and $\bm{u}_f$ are given by
    \begin{gather}
        \begin{aligned}
            &\mu\bm{K}_m^{-1}\bm{u}_m+\nabla p_m=\bm{0}\\
            &\mu\bm{K}_f^{-1}\bm{u}_f+\nabla p_f=\bm{0}
        \end{aligned}
        \quad \text{in } \Omega.
    \end{gather}
    \end{subequations}
    \indent A simpler model, called
    \hl{dual-porosity/single-permeability},
    may be obtained formally by setting $\bm{K}_m=\bm{0}$.  It assumes that the
    porous medium acts as storage volume for flow occurring only along
    fractures. {A further simplification, appropriate only for highly connected
    networks of fracture of small extension relative to the domain size,
    consists in the use of a single {equivalent continuum} with upscaled
    properties that account for the combined effect of porous medium and
    fractures.}

    \subsection{Discrete Fracture Models}
    DFMs represent fractures explicitly, modeled as a network of
    (typically planar) surfaces $\Gamma$, immersed in the porous medium. In the fractures
    we typically use a Darcy-type model where some special source terms are added
    to account for the fluid exchange with the porous medium. The Darcy equations in
    the latter are also modified, with terms that act as interface conditions.

    {DFMs are computationally more demanding that CFMs, but also more
    accurate, particularly when fracture of relative large extension are present. For this reason the choice between CFM and DFM may depend on the
    spatial scale of interest, as illustrated in Fig.~\ref{fig:scale}}.

\begin{figure}
         \centerline{\resizebox{0.35\textwidth}{!}{\fontsize{0.7cm}{2cm}\selectfont
\begingroup%
  \makeatletter%
  \providecommand\color[2][]{%
    \errmessage{(Inkscape) Color is used for the text in Inkscape, but the package 'color.sty' is not loaded}%
    \renewcommand\color[2][]{}%
  }%
  \providecommand\transparent[1]{%
    \errmessage{(Inkscape) Transparency is used (non-zero) for the text in Inkscape, but the package 'transparent.sty' is not loaded}%
    \renewcommand\transparent[1]{}%
  }%
  \providecommand\rotatebox[2]{#2}%
  \newcommand*\fsize{\dimexpr\f@size pt\relax}%
  \newcommand*\lineheight[1]{\fontsize{\fsize}{#1\fsize}\selectfont}%
  \ifx\svgwidth\undefined%
    \setlength{\unitlength}{199.69221461bp}%
    \ifx\svgscale\undefined%
      \relax%
    \else%
      \setlength{\unitlength}{\unitlength * \real{\svgscale}}%
    \fi%
  \else%
    \setlength{\unitlength}{\svgwidth}%
  \fi%
  \global\let\svgwidth\undefined%
  \global\let\svgscale\undefined%
  \makeatother%
  \begin{picture}(1,0.51499931)%
    \lineheight{1}%
    \setlength\tabcolsep{0pt}%
    \put(0,0){\includegraphics[width=\unitlength,page=1]{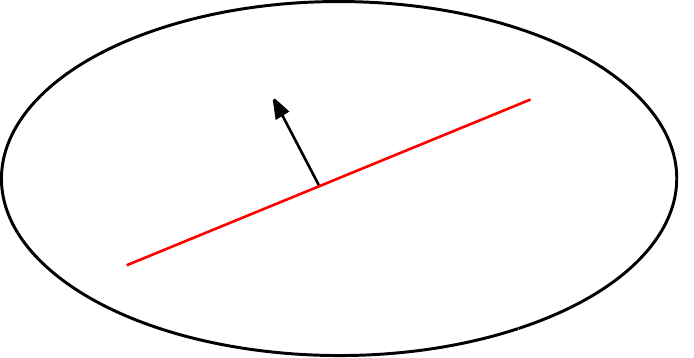}}%
    \put(0.67189847,0.22055065){\color[rgb]{1,0,0}\makebox(0,0)[lt]{\lineheight{1.25}\smash{\begin{tabular}[t]{l}$\Gamma$\end{tabular}}}}%
    \put(0.29375829,0.37908815){\color[rgb]{0,0,0}\makebox(0,0)[lt]{\lineheight{1.25}\smash{\begin{tabular}[t]{l}$n_f$\end{tabular}}}}%
    \put(0.1571775,0.19698049){\color[rgb]{0,0,0}\makebox(0,0)[lt]{\lineheight{1.25}\smash{\begin{tabular}[t]{l}+\end{tabular}}}}%
    \put(0.29186126,0.08695698){\color[rgb]{0,0,0}\makebox(0,0)[lt]{\lineheight{1.25}\smash{\begin{tabular}[t]{l}-\end{tabular}}}}%
  \end{picture}%
\endgroup%
}}
         \caption{Positive and negative side of $\Gamma$.}\label{fig:description}
    \end{figure}
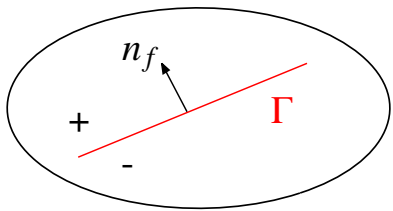

    On each fracture we may identify a unit normal vector $\bm{n}_f$, and thus
    a positive and a negative side of $\Gamma$ ,
    see Fig.~\ref{fig:description}. We  indicate with $[\![f]\!]=f^+-f^-$
    the jump of a quantity $f$
    across the fracture.

    A  commonly used model is derived
    in~\cite{Martin2005}, in which the fracture
    permeability is split into a normal, $K_{f,n}$, and tangential, $\bm{K}_{f,t}$, components
    to account for the fact that those quantities may be different and scale differently
    with the fracture aperture $\epsilon$. In the porous medium
    $\Omega\setminus\Gamma$ we consider Eqs.~\eqref{eq:darcy_cont},
    while in the fractures we have, neglecting the compressibility term,
    \begin{subequations}\label{eq:fracture_model}
    \begin{gather}\label{eq:fracture}
        \begin{aligned}
            &\mu \bm{K}_{f,t}^{-1} \bm{u}_f+\epsilon \nabla_{\tau} p_f=\bm{0} \\
            &\nabla_{\tau}\cdot \bm{u}_f-[\![\bm{u}_m\cdot\bm{n}_f]\!]=q_f
        \end{aligned}\quad \text{in } \Gamma,
    \end{gather}
    where $\nabla_{\tau}\cdot$ and $\nabla_{\tau}$ are the divergence and
    gradient operating on the tangent plane of the fractures, respectively.  The
    jump of normal velocity across the fracture acts as a source term in the
    mass conservation equation and represent the net flux entering (or leaving) the fracture. This model must be complemented by appropriate
    boundary conditions.  In the portion of the boundary of  $\Gamma$ that
    touches the boundary of $\Omega$, the conditions are determined by the
    specific problem at hand: either pressure or mass flux is prescribed.
    In
    the case of a fracture tip that ends inside $\Omega$, a zero mass flux
    condition is generally adopted.

    A network normally exhibits intersection between fractures.
    A common approach is to assume that the pressure is continuous at the
    intersection and the net sum of fluxes is zero. More sophisticated models are
    available, to account for fractures with different hydraulic properties and
    flow along fracture intersections.

    To close the problem we need also to specify
    interface conditions to couple $\Gamma$ to the porous medium. In~\cite{Martin2005}
    a family of models is presented, including the following, often adopted in practice
    \begin{gather}\label{eq:coupling}
        \begin{aligned}
            & \mu \epsilon K_{f, n}^{-1} \bm{u}_m^+ \cdot \bm{n}_f+ 2(p_f
            - p_m^+) = 0\\
            & \mu \epsilon K_{f, n}^{-1} \bm{u}_m^- \cdot \bm{n}_f + 2(p_m^-
            - p_f) = 0
        \end{aligned}
        \quad\text{on } \Gamma.
    \end{gather}
    \end{subequations}
    These conditions can be interpreted as the application of a discrete Darcy law
    across the fracture, {indeed they link the flux across the fracture to pressure differences.}

    \begin{remark}
        If the fractures are highly permeable in the normal direction,
        continuity of pressure across $\Gamma$ is often assumed, i.e.
        $p_m^+=p_m^-$. This induces a certain simplification in the model, since
        it does not account for {a net} mass flow across the fracture, but
        only for porous medium-fracture exchanges.
    \end{remark}

    \begin{remark}
        {For nearly impermeable porous media, a different simplification of these
        models, called \hl{Discrete Fracture Networks} (DFNs), consists in neglecting the effect
        of the porous material and simulate flow just in the fracture network.} They can be used
    in the presence of highly connected and permeable fractures.
    \end{remark}

\section{Discretization schemes for DFM}
Numerical {schemes for the approximation of} the equations presented in the
previous section are based on partitioning the domain $\Omega$ into a mesh
of polyedral elements $\mathcal{T}_h.$ The unknowns are then discretized by
assuming a given variation inside each element, for instance constant or linear.
The continuous solution is then replaced by the discrete values at the mesh nodes.

\begin{figure}[tbp]
    \centering
    \includegraphics[width=0.4\textwidth]{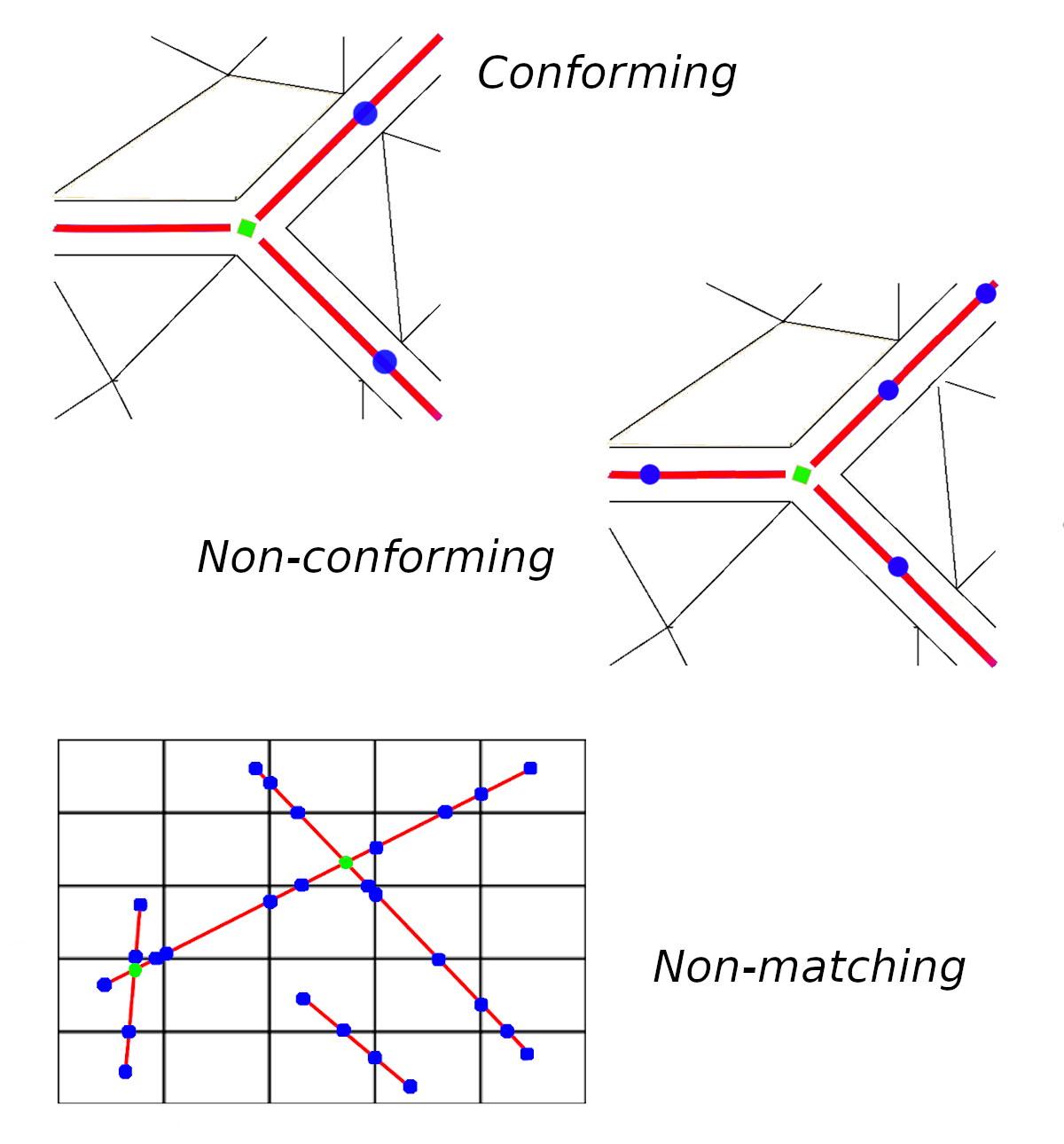}
    \caption{Porous medium and fracture grids.}%
    \label{fig:meshes}
\end{figure}

In the case of DFM we need to mesh both the porous medium and fracture domains and construct suitable
ways to couple the two via \eqref{eq:coupling}. The many discretization
techniques
available in the literature may be roughly subdivided into three categories, depending on the
relation between porous medium and fracture grids,  see Fig.~\ref{fig:meshes}.

\textit{i)} \hl{Conforming methods}. The mesh for
the porous medium conforms to that adopted for the fracture network. It means
that a fracture mesh element {coincide (geometrically) with} the faces of two
porous medium mesh elements, on the positive and negative side of the fracture. Consequently,
no porous medium mesh elements are cut by the fracture. This requirement
poses strong constraints on the mesh generation process, which indeed can be the
most time-consuming part of the simulation, particularly in the presence of
complex networks. On the other hand, the implementation of~\eqref{eq:coupling} is rather straightforward.

\textit{ii)} \hl{Non-conforming methods}. Still no elements in the porous medium grid are cut by the
fractures, however the grid at the two sides of the fracture, as well as the fracture grid, are independent.
The implementation of~\eqref{eq:coupling} involves the set up of suitable operators to transfer the discrete solution in  the fractures
to the porous medium grid and viceversa. The so-called mortar technique, which is based on the set up of additional
variables at the interface, is sometimes used to simplify the construction of the transfer operators.
The process of grid generation is eased and it is simpler to avoid the generation of highly distorted elements,
however it is still rather demanding in complex situations.

\textit{iii)} \hl{Non-matching methods}. The mesh for the porous medium and
the fracture network are completely independent and the fracture grids can cut the porous medium mesh elements in
an arbitrary way. This simplifies mesh generation greatly, since porous medium and fracture grids can be generated independently.
It is still necessary to find the intersections, but this is simpler than generating a conforming mesh and can be done with standard geometric search tools.
 However, the implementation of conditions~\eqref{eq:coupling} is more complex.

For every category, many different \hl{numerical schemes} are at disposal. For the
cases \textit{i)} and \textit{ii)}, it is beneficial to use techniques able to operate on
arbitrary polyhedral grids, like Finite Volumes or Mimetic Finite Differences,
to mention the more established one. The research in this field is, however,
very active, and we mention also Gradient Schemes, the Hybrid High Order (HHO)
method and the Virtual Elements Method (VEM).  A reference containing examples
of numerical schemes applied in a DFM context is~\cite{Fumagalli2019a}.

In the case \textit{iii)}, as already stated, the main difficulty is how to impose the
interface conditions. In that respect we have two class of procedures. The first
is to represent the possible jumps in the solution across the elements cut by a
fracture explicitly. This is what is done in eXtended Finite Elements (XFEM),
where the finite element basis functions are locally enriched to allow
discontinuous solutions that can satisfy the coupling conditions
~\eqref{eq:coupling}. Contrarily, if one accepts a less accurate representation
of the solution, some manipulations of the interface conditions are possible to
transform them into source terms acting both on the fracture elements and on the
elements in the
porous medium that are cut by the fracture. The resulting source terms have
some similarity with those present in CFM techniques. The Embedded Discrete
Fracture Model (EDFM), usually coupled with a simple Finite Volumes
approximation, falls into this second category of methods. An reference on EDFM
techniques is in~\cite{Li}.

The performances of many classes of methods are presented and discussed in
\cite{Flemisch2016a} for bi-dimensional problems and \cite{Berre2020} for three-dimensional problems.

\section{An example of computational workflow}\label{sec:example}

Numerical simulation of flow in fractured porous media is challenging due to the
intrinsic geometrical complexity of the fractures, {as well as the measurement
of real fractures and their properties buried deep in the underground.  These
data are difficult to obtain and usually affected by large uncertainty which
compromises the reliability of the numerical solutions.

Several approaches can be considered to detect fractures in the underground,
from seismic inversion to  \hl{outcrop interpretation}, the former effective to
detect big fractures few kilometres below the surface while the latter normally
used as an analogue of the underground. Once the fractures are collected and
digitalized by means of one of these methods, a suitable mathematical model can
be adopted to perform the simulations.

We focus our attention on the case of outcrop interpretation.  From highly
detailed photographs of the interested region, all the fractures are collected
and interpreted  up to a minimum sampling scale determined by the quality of the
data or by computational constraints.  Smaller fractures can be {accounted for
by a suitable change of the } porous medium properties and seen as upscaled or
homogenized. What quantity defines a fracture ``small'' and thus not explicitly
represented is still an open question, many authors consider the fracture
extension a good proxy for its importance. After this operation, the digitalized
version of the outcrop is thus available, see Fig.~\ref{fig:sotra} on the top
as an example where natural and human factors limit its exposure.  The outcrop
is a portion of Sotra island, near Bergen in Norway.

Another challenging aspect is the collection of the physical data, again
affected by uncertainty, that are needed to run the simulation. If some data are
not available, it is possible to fill the gap  considering models from the
literature to, for example, relate the permeability with the aperture and the
latter with the fracture extension. {The model can, at this point, be
discretized numerically by means of one of the methods mentioned in the previous
sections.}

Fig.~\ref{fig:sotra} reports numerical results of a simulation, in particular
the pressure and concentration of a tracer (for instance a contaminant) in the
interpreted outcrop. Two extreme cases are considered, if the fractures are
higher or lower permeable than the surrounding porous medium. Data are homogeneous
and a left to right pressure gradient is imposed. The solutions are obtained
with the library PorePy, see \url{https://github.com/pmgbergen/porepy}.

\begin{figure*}[tbp]
    \centering
    \subfloat[Interpreted outcrop.]%
        {\includegraphics[width=0.475\textwidth]{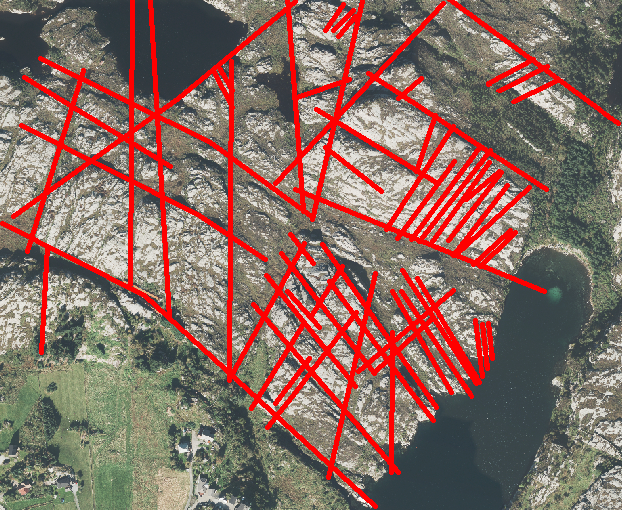}}%
    \hfill%
    \subfloat[Digitalized fractures.]%
        {\includegraphics[width=0.475\textwidth]{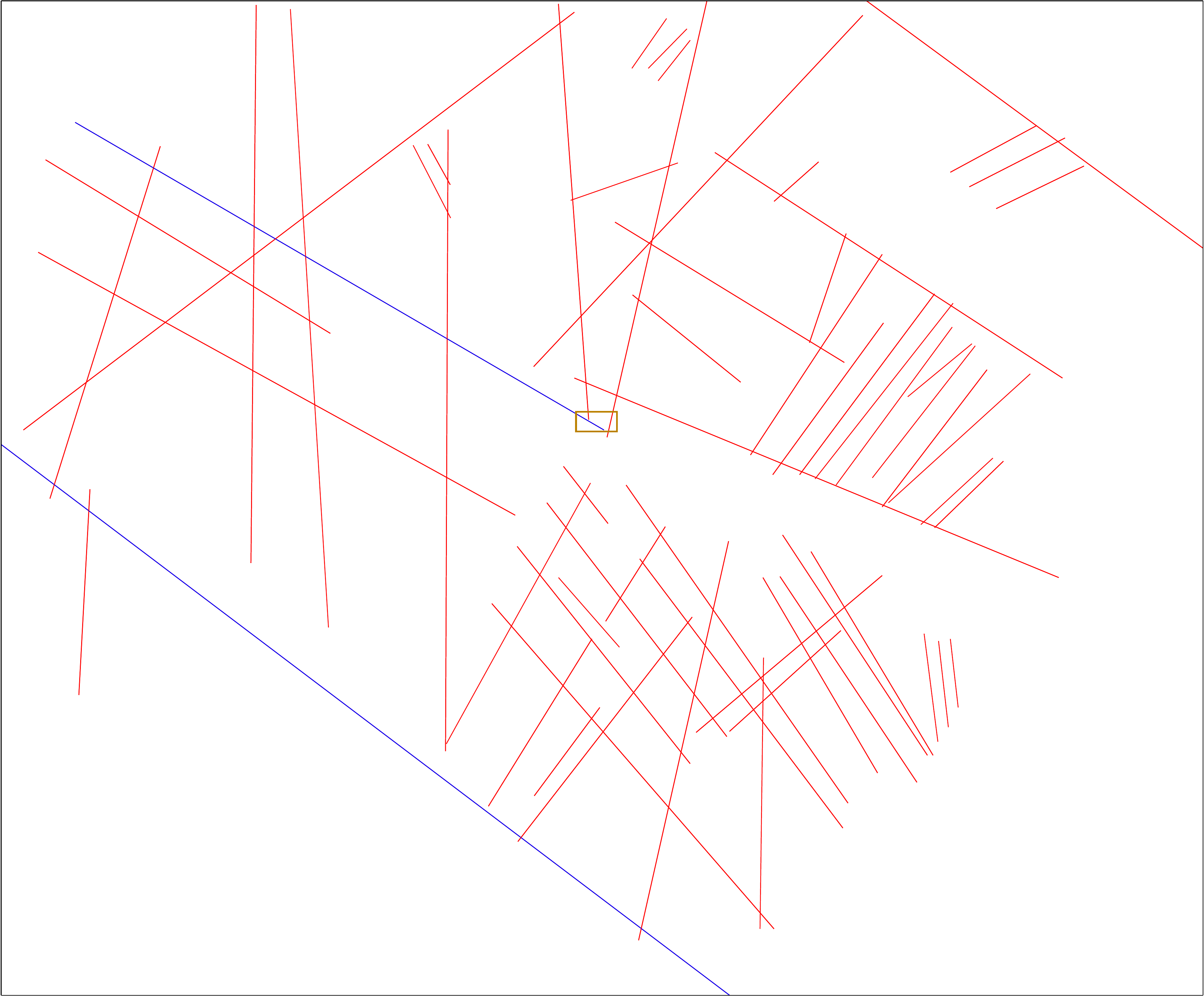}}\\
    \subfloat[$p$ and $\bm{u}$ with high permeable fractures.]%
        {\includegraphics[width=0.475\textwidth]{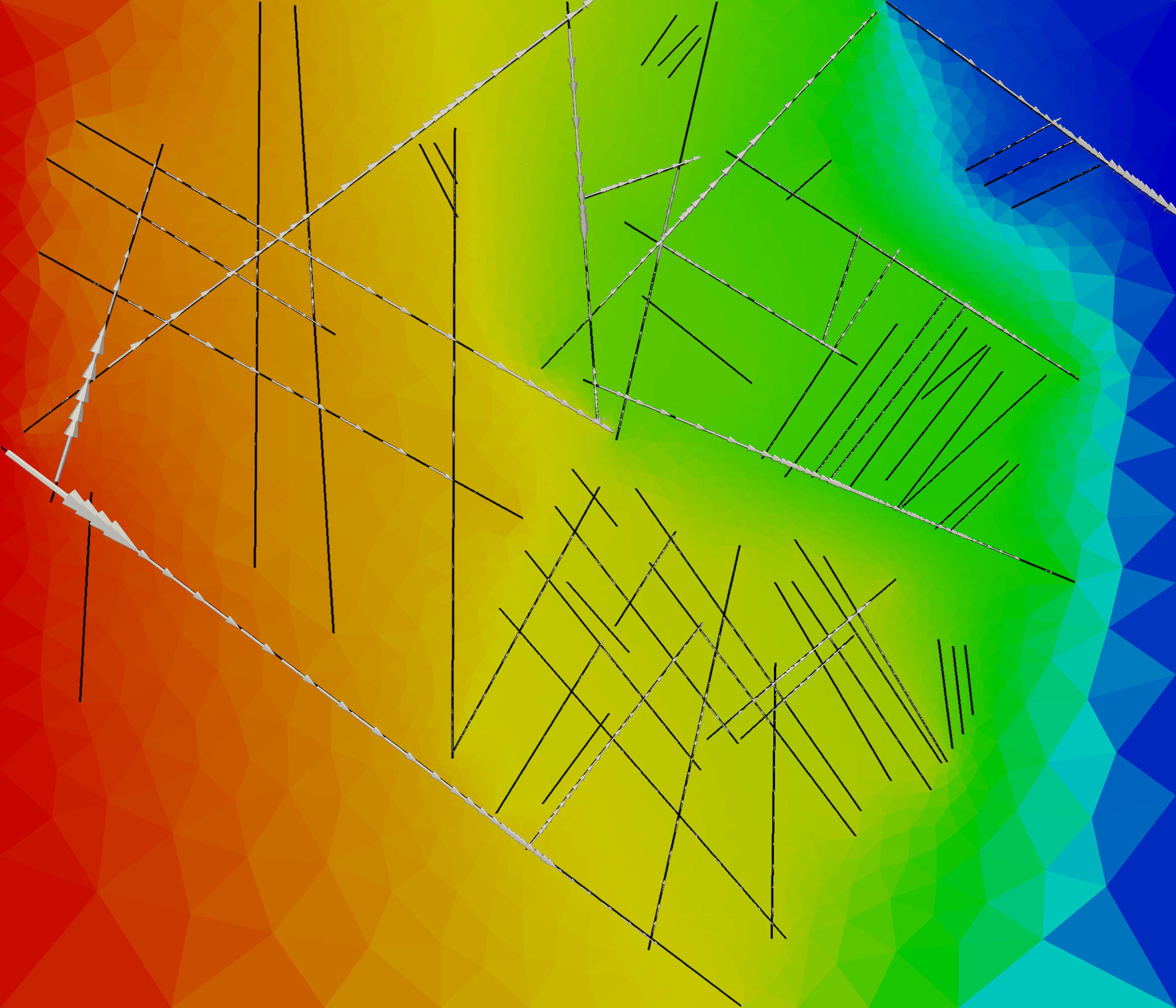}}%
    \hfill%
    \subfloat[Scalar with high permeable fractures.]%
        {\includegraphics[width=0.475\textwidth]{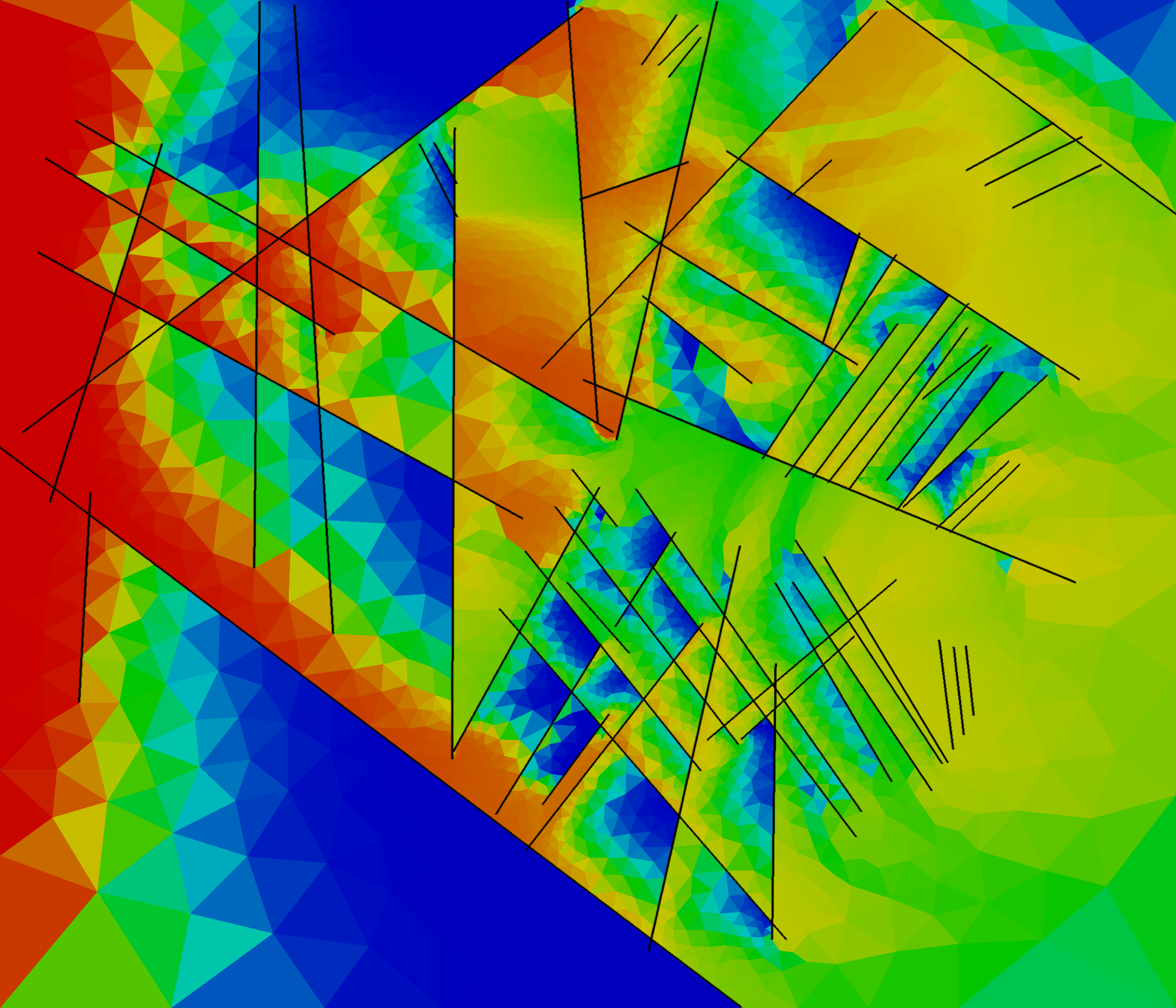}}\\
    \subfloat[$p$ and $\bm{u}$ with low permeable fractures.]%
        {\includegraphics[width=0.475\textwidth]{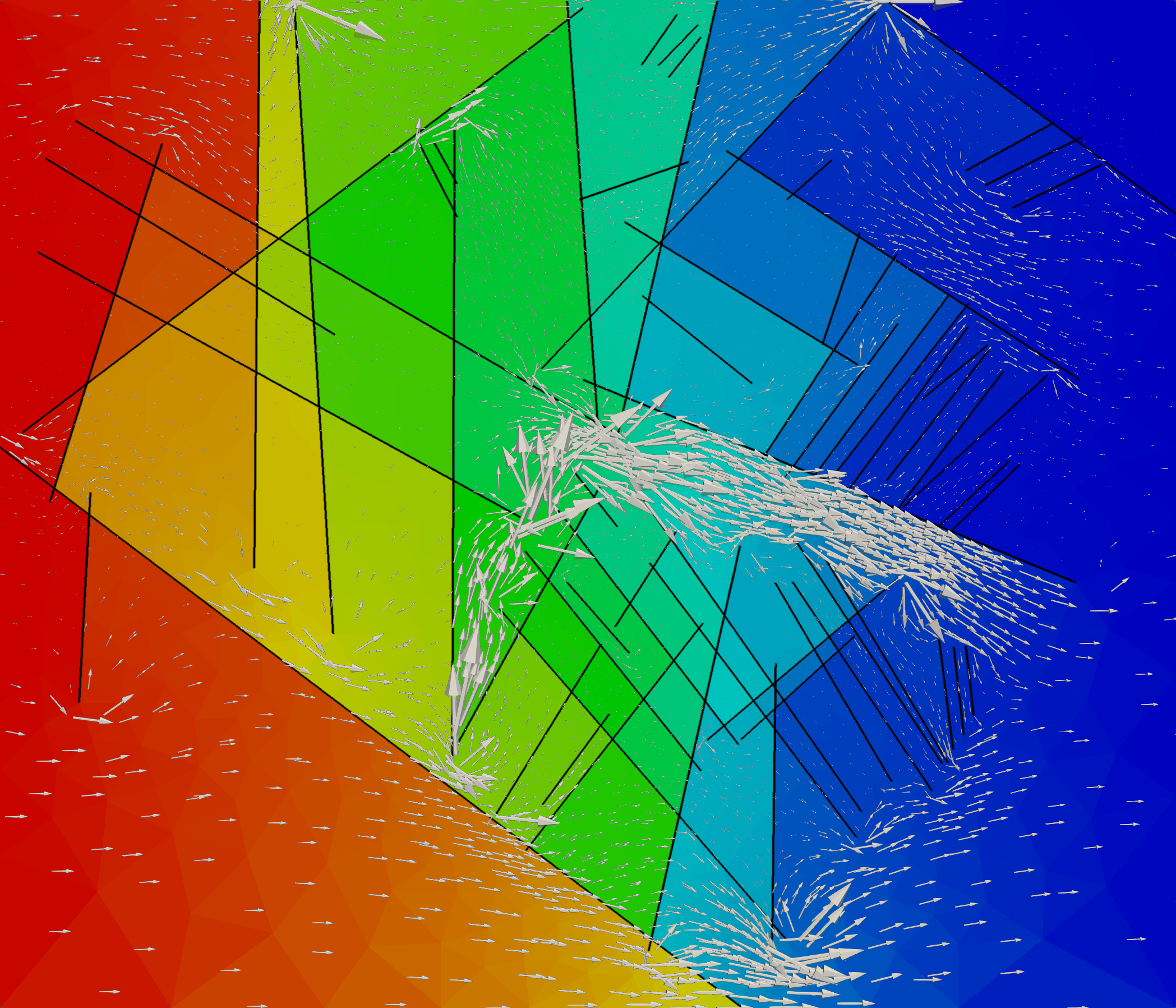}}%
    \hfill%
    \subfloat[Scalar with low permeable fractures.]%
        {\includegraphics[width=0.475\textwidth]{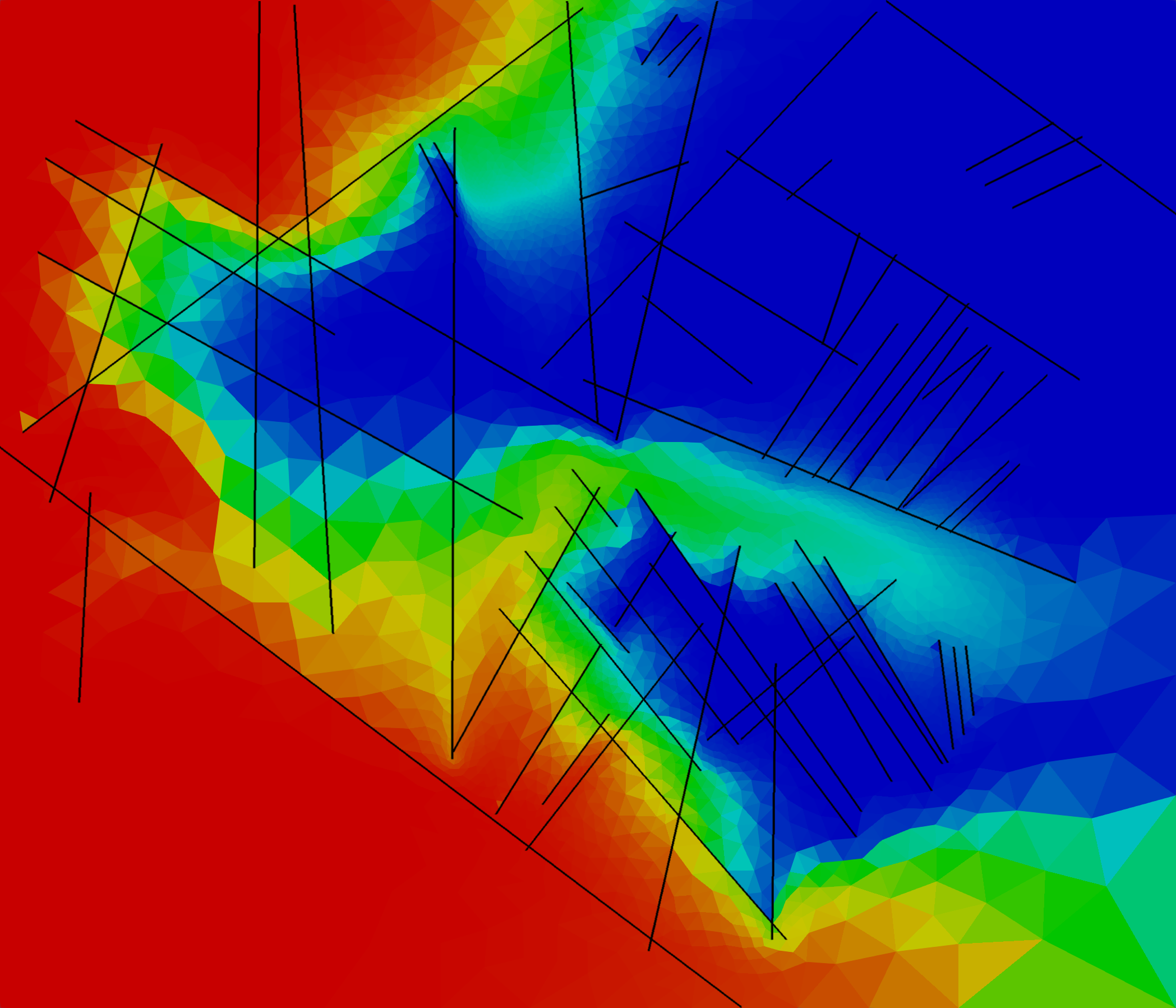}}
    \caption{On the top the interpreted outcrop and digitalized fractures.
        The blue fractures are geometrically simplified due to software
        constraint.  Images are taken from \cite{Flemisch2016a}.
        The others represent pressure, velocity field, and a scalar
        tracer at a specific time. The colour
        scheme spans from the lowest in blue to the highest in red. Images are taken
        from Fumagalli, A., Keilegavlen, E.: Dual virtual element methods for discrete
        fracture matrix models.
        \newblock Oil \& Gas Science and Technology - Revue d'IFP Energies nouvelles
        {74}(41), 1--17 (2019). under CC BY 4.0 licence.}%
    \label{fig:sotra}
\end{figure*}

\section{Conclusion}
Modeling flow in porous media in presence of fractures and fracture networks is
a challenging task. Several approaches are available in literature, and a few of
them implemented in specialised software. The choice depends on the scales at
which the phenomenon has to be considered, on the connectivity of the fracture
network and on the level of accuracy desired. Simpler continuum fracture models
are suitable for highly connected and dense networks of fracture, while in
presence of fracture with a larger extension and more sparse, discrete fracture models
provide more accurate results. Clearly, it is also possible to use a combination of the two approaches. In this chapter we gave a rapid review of the different
strategies and, for the sake of simplicity, we focused the attention on
single-phase flow. The  general conclusions can however be extended to the more
complex situation of multi-phase flows.

\begin{acknowledgement}
The authors acknowledge Davide Losapio for his help in the research underlying this manuscript.
\end{acknowledgement}

\section*{Cross-References}

\begin{itemize}
    \item Lithosphere, Mechanical Properties
    \item Numerical Methods, Finite Element
    \item Sedimentary Basins
\end{itemize}

%
%


\end{document}